\documentclass[a4paper]{article}

\usepackage{INTERSPEECH2022}

\usepackage{cite}
\usepackage{url}
\usepackage{hyperref}
\hypersetup{
     colorlinks=true,
     citecolor=blue,
}
\usepackage{multirow}
\usepackage{makecell}
\usepackage{booktabs}

\usepackage{etoolbox}
\newtoggle{arxiv}
\toggletrue{arxiv} 

\title{The VoiceMOS Challenge 2022}
\name{
Wen-Chin Huang$^{1\dagger}$\thanks{$^{\dagger}$Equal contribution.}, Erica Cooper$^{2\dagger}$, Yu Tsao$^3$, Hsin-Min Wang$^3$, Tomoki Toda$^1$, Junichi Yamagishi$^2$
}
\address{
  $^1$Nagoya University, Japan\\
  $^2$National Institute of Informatics, Japan\\
  $^3$Academia Sinica, Taiwan
}
\email{voicemos2022@nii.ac.jp}

\begin{document}

\maketitle
\begin{abstract}
We present the first edition of the VoiceMOS Challenge, a scientific event that aims to promote the study of automatic prediction of the mean opinion score (MOS) of synthetic speech.  This challenge drew 22 participating teams from academia and industry who tried a variety of approaches to tackle the problem of predicting human ratings of synthesized speech.  The listening test data for the main track of the challenge consisted of samples from 187 different text-to-speech and voice conversion systems spanning over a decade of research, and the out-of-domain track consisted of data from more recent systems rated in a separate listening test.  Results of the challenge show  the effectiveness of fine-tuning self-supervised speech models for the MOS prediction task, as well as the difficulty of predicting MOS ratings for unseen speakers and listeners, and for unseen systems in the out-of-domain setting.
\end{abstract}

\section{Introduction}

Speech synthesis technologies such as text-to-speech synthesis (TTS) and voice conversion (VC) are a very active area of research, and this field relies on judgments of quality of the synthesized speech.  The gold standard for evaluation is listening tests, where human raters listen to samples generated by different synthesis methods and give their opinions.  One popular type of listening test is Mean Opinion Score (MOS) \cite{recommendation2006vocabulary}, in which listeners are presented with synthesized samples one by one, and asked to rate some aspect of the speech, such as how natural it sounds, on a Likert scale, typically from 1-5.  

With the popularity of crowdsourcing platforms, conducting listening tests has become easier than in the past, when researchers would have to recruit and schedule participants locally to come into the lab to listen to samples in person.  MOS tests also enable the implicit comparison of many different synthesis systems -- tests such as A/B testing, which ask listeners to directly compare samples from two different systems, quickly become impractical as the number of systems increases, since all pairs of systems must be compared.  Nevertheless, listening tests in general and MOS tests in particular are not without their drawbacks.  Listening tests are time-consuming and costly, and they cannot be used as a loss term in a training objective when developing synthesis systems.  Furthermore, listening tests are heavily context-dependent.  Each listening test gathers opinions from a different set of listeners, and asks them to consider different sets of systems, whose overall range of quality may be very different.  The lexical content of the synthesized samples and even the instructions given to participants also differ from test to test.  This means that although MOS testing results in numerical values representing the quality of each system, these numbers cannot be meaningfully compared across different listening tests.

Intrusive speech quality metrics such as PESQ \cite{pesq} are not suitable for evaluating synthesized speech because they require matching ground-truth reference audio. Synthesized speech may have different prosody from a ground-truth reference, unlike natural speech processed by a codec or in noisy conditions, for which these metrics were designed.  Given that MOS tests result in a  set of audio samples paired with their ratings from human listeners, there is no surprise that researchers have begun to apply data-driven, machine-learning-based approaches to develop automatic non-intrusive MOS prediction models \cite{moller2010comparison, linear-reg-BC2012, norrenbrock2015quality, Yoshimura2016, patton2016automos, mosnet, siamese-mos, Choi2020, williams2020comparison, mbnet, s3r-mos}. However, given the context-dependency of MOS ratings, data from different listening tests cannot be combined, and so these models are typically trained on data from one listening test that contains samples from a limited range of types of synthesis models.  While these MOS prediction models can perform well for samples from the same listening test, they typically fail to generalize well to other listening test contexts or even to unseen synthesis methods from the same listening test \cite{generalization-mos}.  

In order for MOS prediction models to be a useful tool for speech synthesis researchers, these issues need to be addressed and more improvement is needed in the state of the art in the MOS prediction task.  With these aims, we developed the \textbf{VoiceMOS Challenge,} a shared task using common datasets for MOS prediction.  We conducted a large-scale MOS listening test to cover many different types of TTS and VC systems spanning many years \cite{BVCC, generalization-mos}, and this dataset forms the basis of the challenge.  We also introduced an ``out-of-domain" track to focus research efforts on how to adapt MOS prediction models to new listening test contexts. In the following, we describe the organization of the challenge and present some analysis of the predictions made by participants' MOS prediction systems.

\section{Challenge Description}

The challenge was hosted on CodaLab\footnote{\url{https://codalab.lisn.upsaclay.fr/competitions/695}}, an open-source web-based platform for reproducible machine learning research. Below we introduce the dataset used in each track. A summary is in Table~\ref{tab:data}. We also describe the challenge phases, evaluation metrics, and baseline prediction systems.
\iftoggle{arxiv}{
    The challenge rules can be found in Appendix~\ref{appendix:rules}.
}

\subsection{Tracks and Datasets}

\subsubsection{Main track}

The data for the main track of the challenge comes from a large-scale listening test covering many types of TTS and VC systems that we conducted in our prior work \cite{BVCC}.  The listening test consists of 38 samples from each of 187 different synthesis systems including natural speech, and each sample was rated by 8 different listeners.  The samples come from many past years of Blizzard Challenges (BC) \cite{blizzard2008,blizzard2009,blizzard2010,blizzard2011,blizzard2013,blizzard2016} and Voice Conversion Challenges (VCC) \cite{vcc2016,vcc2018,vcc2020}, as well as published samples from ESPnet-TTS \cite{hayashi2019espnettts}.  These samples cover a wide range of different types of synthesis systems and methods spanning many years.  Listeners rated each sample for naturalness on a discrete, 1-incremented scale from 1 (very bad) to 5 (very good).  We also collected demographics about listeners such as gender, age range, and whether or not they have any hearing impairment.  We used the training, development, and test splits from our prior work \cite{generalization-mos} which followed a 70\%/15\%/15\% partition, and which were chosen to hold out some unseen synthesis systems, speakers, texts, and listeners in the development and test sets while matching the overall distributions of ratings as closely as possible between the sets.

\subsubsection{Out-of-domain track}

We define ``out-of-domain'' (OOD) as data that is from a different listening test.  For our OOD track, we chose the Blizzard Challenge 2019 listening test data \cite{wu2019blizzard}, which was not included in the main track data.  In addition to this data originating from a different listening test with different synthesis systems and listeners from the main track, this dataset consists of Chinese TTS samples whereas all of the samples in the main track dataset are in English, which presents a challenging language mismatch in the domain as well.  The OOD data was split into training/unlabeled/development/test sets of 10\%/40\%/10\%/40\%.  The unlabeled dataset was audio samples only without MOS ratings, provided to participants during the training phase in order to encourage experimentation with approaches such as semi-supervised learning.  The small training set size was chosen to reflect a scenario where a small amount of labeled target-domain data is available, e.g.,\ from a small pilot listening test.  The main track data was allowed to be used in the OOD track in addition to the OOD data.  We have made the scripts for obtaining the challenge data publicly available.\footnote{\url{https://doi.org/10.5281/zenodo.6572573}}

\begin{table}[t]
	\centering
	\caption{Summary of the main track and out-of-domain (OOD) track datasets.}
	
	\centering
	\begin{tabular}{ c c c c c c}
		\toprule
		\multirow{2}{*}[-1pt]{Track} & \multirow{2}{*}[-1pt]{Lang} & \multicolumn{3}{c}{\# Samples} & \multirow{2}{*}[-2pt]{\makecell{\# ratings\\per sample}}\\
		\cmidrule(lr){3-5}
		& & Train & Dev & Test & \\
		\midrule
		Main & Eng & 4,974 & 1,066 & 1,066 & 8 \\
		\midrule
		OOD & Chi & \makecell{Label: 136\\Unlabel: 540} & 136 & 540 & 10-17\\
		\bottomrule
	\end{tabular}
	\label{tab:data}
\end{table}

\subsection{Phases}

The challenge was divided into four phases: the training phase, evaluation phase, break phase, and post-challenge.  

\subsubsection{Training phase}

The training phase started on December 5, 2021, and ended on February 21, 2022.  During these 11 weeks, participants had access to audio samples and their MOS labels and listener info for the training and development data (as well as the unlabeled audio set for the OOD track), and the purpose of this phase was for participants to develop their prediction systems.  Participants could submit their MOS predictions for the development set to the CodaLab leaderboard, where teams' scores were publicly displayed.  Participants were permitted to make up to 30  submissions to the leaderboard per day.

\subsubsection{Evaluation phase}

The evaluation phase was one week long, starting on February 21, 2022 and ending on February 28.  Test set audio samples were released at the start of this phase. Participants did not have access to any MOS labels or listener info for the test set.  Teams were permitted to make up to three submissions during this phase with their MOS predictions for the test set audio.  

\subsubsection{Break phase and post-challenge phase}

We defined a 1-week break phase after the end of the evaluation phase, during which we froze new submissions so that we could conduct post-challenge analysis.  After that was the post-challenge phase, during which we re-opened the leaderboard so that participants could continue making submissions.

\subsection{Evaluation Metrics}

Following \cite{mosnet}, the evaluation criteria are system-level and utterance-level mean squared error (MSE), Linear Correlation Coefficient (LCC), Spearman Rank Correlation Coefficient (SRCC), and Kendall Tau Rank Correlation (KTAU). The reason to use multiple metrics is to assess the model performance via various aspects. Depending on the application, different metrics have their own respective roles. In scientific challenges like BC or VCC, we are more interested in the ranking of synthesis systems, so metrics like SRCC or KTAU are preferable. On the other hand, when it comes to assessing TTS/VC models under development, MSE is a more straightforward metric.

We distributed an evaluation script to avoid confusion and inconsistency. Due to the space limit, we only displayed utterance-/system-level MSE/SRCC on the leaderboard. Following \cite{s3r-mos}, we used system-level SRCC as the primary metric for determining the ranking on the leaderboard.

\subsection{Baseline Prediction Systems}

We provided three open-source baseline MOS prediction systems to participants: SSL-MOS\footnote{\url{https://github.com/nii-yamagishilab/mos-finetune-ssl}} \cite{generalization-mos}, MOSA-Net\footnote{\url{https://github.com/dhimasryan/MOSA-Net-Cross-Domain}} \cite{mosanet}, and  LDNet\footnote{\url{https://github.com/unilight/LDNet}} \cite{ldnet},  which correspond to team IDs B01, B02 and B03,
respectively. SSL-MOS adds a simple linear fine-tuning layer for the MOS prediction task onto Fairseq\footnote{\url{https://github.com/pytorch/fairseq}} self supervised learning  (SSL)-based models for speech, and in particular, the baseline is fine-tuned from Wav2vec 2.0 Base \cite{wav2vec2}, which was trained on the LibriSpeech \cite{LibriSpeech} corpus.  MOSA-Net uses cross-domain features such as spectral information, complex features, raw waveform data, and features extracted from SSL models to estimate multiple speech assessment metrics simultaneously. Originally developed for noisy speech quality assessment, MOSA-Net serves as a good example to demonstrate how adapting such a prediction system to this task is possible.  LDNet does not rely on any external data, but its specialized model structure and inference method allow for utilizing individual ratings to facilitate listener-dependent modeling. 

\section{Participants and Submitted Systems}

In total, 22 teams made an evaluation phase submission, where 14 are from academia, five are from industry, two are joint academia-industry teams, and one is personal. Participants span from countries in Asia such as Japan, Taiwan, and China, to those in Europe, including the UK, Ireland, the Netherlands, Hungary, Romania, and Czechia. Among them, 21 and 15 teams participated in the main and OOD tracks, respectively. 
\iftoggle{arxiv}{
    A full list of participants and their affiliations can be found in Appendix~\ref{appendix:participants}.
}

During the training phase, a total of 223 submissions were made, as some teams frequently uploaded results to compete with each other. One of the teams, T17, made 33 submissions and turned out to be one of the top teams.

\section{Results, Discussion and Analysis}

\begin{figure*}[t!]
\centering
\includegraphics[width=1.\linewidth]{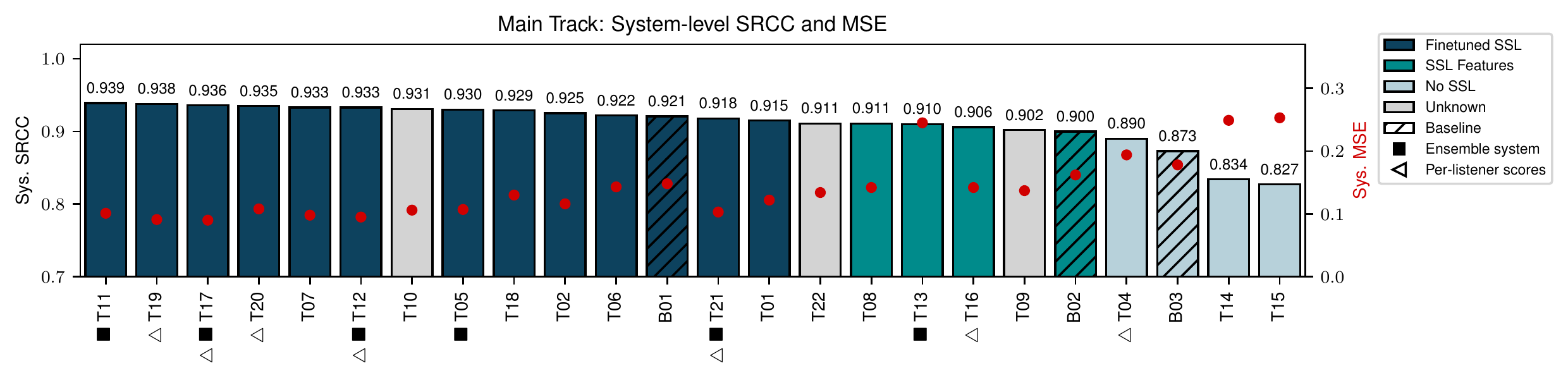}
\vspace{-7mm}
\caption{Main track results}
\label{fig:main}

\end{figure*}

The evaluation results of the main and OOD tracks are shown in Figures \ref{fig:main} and \ref{fig:ood}.
\iftoggle{arxiv}{
    Full numerical results and rankings for all metrics can be found in Appendix~\ref{appendix:detailed-results}. 
} {
    Full numerical results for all metrics can be found in our supplementary materials online\footnote{\url{https://arxiv.org/abs/2203.11389}}.
}

\subsection{Comparison of Baseline Systems}

Among the three baselines, B01 was the strongest in terms of system-level MSE and SRCC for the main track and system-level SRCC for the OOD track. B02 topped the system level MSE for the OOD track, and placed second among the other metrics. B03 was consistently the worst among the three baselines. As both B01 and B02 used SSL, it is shown that using extra data is helpful even if MOS ratings are absent.

For the main track, B01 had a system-level MSE and SRCC of 0.148 and 0.921, ranking 18th and 12th in the two metrics, respectively. The top prediction systems in system-level MSE and SRCC scored 0.090 and 0.939, respectively. As for the OOD track, B02 had a system-level MSE of 0.071, and B01 had a system-level SRCC of 0.975. The top prediction systems had an MSE of 0.030 and a SRCC of 0.979, respectively. This suggests that the gap between the baseline and the top prediction systems was not large. As one of the points of feedback from the participants suggests, the baseline was already strong and it was hard to obtain substantial improvements over it.

\vspace{-2mm}

\begin{figure*}[t!] 
\centering
\includegraphics[width=1.\linewidth]{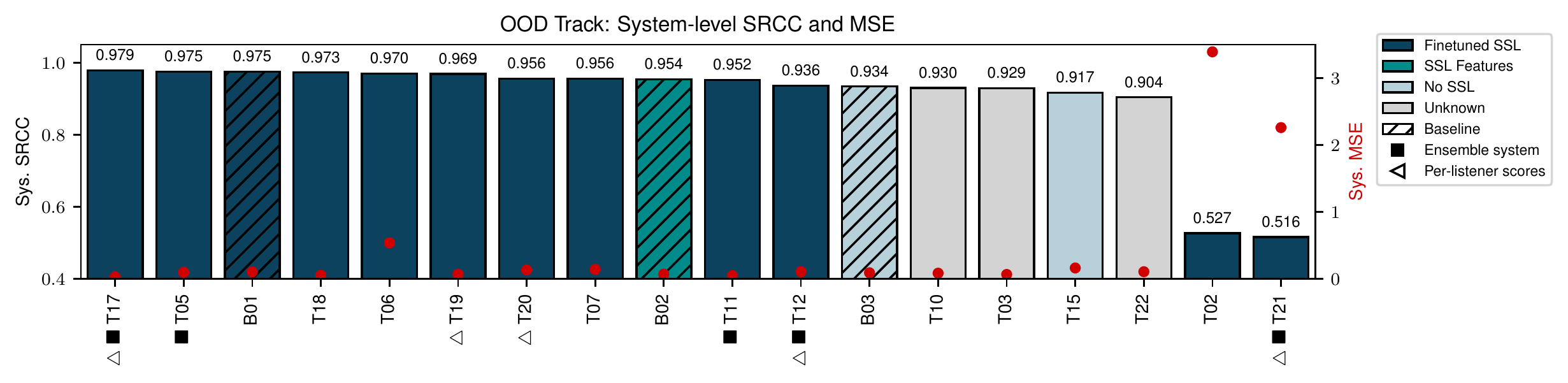}
\vspace{-7mm}
\caption{OOD track results}
\label{fig:ood}

\end{figure*}

\subsection{Analysis of Top Prediction Systems}

The top-ranking teams in the main track all use methods that involve fine-tuning SSL models (either using our SSL-MOS baseline code or not).  In fact, for the main track we can see from Figure \ref{fig:main} that teams are clearly arranged with SSL-finetuning-based approaches scoring the highest, models that use features extracted from SSL models without finetuning in the middle, and methods that do not use SSL ranking last. However, for the OOD track this arrangement does not hold true, with some teams that did not finetune SSL models outranking ones that did. Some popular modeling approaches that we observed in both the top prediction systems and the teams overall are ensembling, multi-task learning, and use of speech recognizers.

Seven teams (and 3 of the top 5 teams) made use of per-listener ratings, either by using the LDNet baseline or by implementing a similar method.  No teams made use of the listener demographics that were provided, but one team made use of ``listener group,'' since the structure of the main-track listening test had listeners grouped by the set of utterances that they rated.

For the OOD track, only three teams tried making use of the unlabeled data.  T17  conducted their own listening test to get labels for the unlabeled data, and this strategy was very successful as they ranked first for system-level SRCC in the OOD track.  Other strategies were to use the unlabeled data for task-adaptive pretraining \cite{gururangan-etal-2020-dont}, and to use trained models to ``self-label'' the unlabeled data.  System-level SRCCs were overall higher in the OOD phase than in the main phase, but the MSEs also spanned a wider range; this indicates that although relative rankings of synthesis systems can be well-learned by pretraining on a larger dataset, predicting the precise scores for a different listening test with very little data from that test remains a challenge.

\subsection{Sources of Difficulty for MOS Prediction}

\subsubsection{Seen vs. unseen categories}

Similar to our prior work \cite{generalization-mos}, we analyzed the difficulty of seen vs.\ unseen categories such as synthesis systems, speakers, and listeners.  For each team, we gathered the squared errors for all utterances from e.g.\ seen speakers, and then for unseen speakers, and then conducted a two-sided t-test between these lists of squared errors to determine whether the unseen-category errors were significantly different (either more difficult or less difficult) from the seen ones at a level of p$\leq$0.05.  

Surprisingly, for the main track, we found that no teams found unseen synthesis systems to be significantly more difficult by this measure, except for B01 (as reported in \cite{generalization-mos}).  Unseen listeners were significantly more difficult for 16 teams and 1 baseline (and significantly easier for 1 team).  There was only one unseen speaker in the test set, but unseen speaker data was significantly more difficult for 6 teams and 1 baseline.

For the OOD track, unlike the main track, unseen synthesis systems were more difficult than seen ones in the OOD test set for 4 teams and 2 baselines.  Unseen listeners were not significantly harder to predict for any teams.  There are no unseen speakers in the OOD track test set.

\subsubsection{Synthesis systems that are difficult to predict}

Since unseen synthesis systems were not especially difficult to predict in the main track, we next tried to identify whether there were any particular synthesis systems that are consistently difficult for all teams, and to identify what makes them difficult.  For all teams including baselines, we ranked synthesis systems based on system-level MSE, and found each team's ``worst-10'' synthesis systems.  Then, for each synthesis system that appears in {\em any} ``worst-10'' list, we count how many teams have that system in their worst-10 list.  The top 5 most difficult synthesis systems, judged as those that appeared in the most teams' worst-10 lists, are in Table \ref{tab:difficultsystems}.

\begin{table}[ht!]
	\centering
	
	\caption{Top 5 most difficult synthesis systems to predict in the main track, and information about them: System name, how many teams had it in their worst-10 list, ground-truth MOS, and standard deviation of MOS.}
	\label{tab:difficultsystems}
	\centering
	\begin{tabular}{ | l | l  l  l | }
	\hline
	Name & \# Teams & MOS & Sdev. MOS \\
	\hline
	BC2009-T & 22 & 2.65 & 0.88 \\
	BC2016-F & 19 & 2.52 & 1.00 \\
	BC2008-B & 15 & 3.22 & 0.98 \\
	BC2008-I & 13 & 2.40 & 1.02 \\
	VCC2020-T10 & 12 & 3.88 & 0.83 \\
	\hline
	
	\end{tabular}
\end{table}

We did not find that factors such as type of synthesis or standard deviations of synthesis systems' ratings could explain their difficulty.  We also considered that differences in the distributions of scores for each synthesis system's training and testing splits may affect difficulty.  Although our training, development, and testing splits were carefully chosen to have well-matched distributions of MOS ratings, this was done for the overall data and not on a finer-grained system-by-system level.  We measured the difference in the distributions of the splits for each synthesis system in the following manner:  For every synthesis system in our main track data, we computed the earth-mover's distance (EMD) between the distribution of the ratings in the training set compared to the test set.  System VCC2020-T10 had no samples in the training set since it was held out as an unseen system for the development set, so it was excluded from this analysis.  We found that all four of the remaining difficult systems fall within the 15\% of systems with the highest EMDs.  Thus, unsurprisingly, we can conclude that a discrepancy in the training and test set distributions can adversely affect prediction ability; this is a well-understood phenomenon in machine learning.

\subsection{Analysis of Metrics}

\vspace{-3mm}

\begin{table}[t]
\centering
\caption{Linear correlation coefficients between system-level metrics, using the main track results.}
\label{tab:metrics-correlations-sys}

\begin{tabular}{ | c | c c c c | }
        \hline
        & MSE & LCC & SRCC & KTAU \\
        \hline \hline
        MSE  & 1.00 & -.875 & -.862 & -.870 \\
        LCC  & - & 1.00 & .997 & .994\\
        SRCC  & - & - & 1.00 & .994\\
        KTAU  & - & - & - & 1.00\\

\hline
\end{tabular}
\end{table}

We observed that ranking top at one metric does not guarantee superiority at other ones. For example, T11 ranked top in system-level LCC, SRCC and KTAU in the main track, but ranked 5th in system-level MSE. To understand the tendencies of each team at each metric, we calculated the linear correlation coefficients between all pairs of metrics using the main track results, as shown in Table~\ref{tab:metrics-correlations-sys}.
\iftoggle{arxiv}{
    A full table of the correlation coefficients of all metrics in both utterance- and system-levels can be found in Appendix~\ref{appendix:metrics-correlations-all}.
}

First, LCC, SRCC and KTAU have coefficients close to 1 with each other in both utterance- and system-levels. However, MSE behaves differently from the above 3 metrics, as evidenced by the relatively lower correlation ($\sim0.87$). This analysis suggests that future studies may choose to report, for example, only the SRCC, while still keeping the MSE. It is also of interest to develop a general metric that takes both metrics into account.

\section{Conclusions}

We have presented a description of the first VoiceMOS Challenge and a summary of the results.  We have observed the overwhelming effectiveness of SSL-based models for this task.  We have also observed the difficulty of unseen synthesis systems in the OOD setting, and of unseen speakers and listeners in the main track.  Generalizing to a different listening test context and predicting scores precisely using a small amount of labeled data remains challenging, as shown by the wide range of MSE values in the OOD track.  We have also observed the importance of well-matched training and test distributions.  Our future challenge will focus on the high-scoring region of the MOS range.  Given the high quality of present-day speech synthesis systems, this task is very relevant to speech synthesis researchers, especially if MOS predictors are to be useful for objectively evaluating experimental synthesis systems.

\vspace{1mm}
\noindent
\footnotesize{\textbf{Acknowledgements} This work was supported by JSPS KAKENHI Grant Number 21J20920, JST CREST Grant Number JPMJCR19A3, JPMJCR18A6 and JPMJCR20D3, and by MEXT KAKENHI grants 21K11951 and 21K19808. This work was also supported by MOST 107-2221-E-001-008-MY3.  We would like to thank the organizers of the Blizzard Challenge and the Voice Conversion Challenge as well as the creators of ESPnet-TTS for making their audio samples available.}

\bibliographystyle{IEEEtran}
\bibliography{mybib}

\begin{thebibliography}{10}
\providecommand{\url}[1]{#1}
\csname url@samestyle\endcsname
\providecommand{\newblock}{\relax}
\providecommand{\bibinfo}[2]{#2}
\providecommand{\BIBentrySTDinterwordspacing}{\spaceskip=0pt\relax}
\providecommand{\BIBentryALTinterwordstretchfactor}{4}
\providecommand{\BIBentryALTinterwordspacing}{\spaceskip=\fontdimen2\font plus
\BIBentryALTinterwordstretchfactor\fontdimen3\font minus
  \fontdimen4\font\relax}
\providecommand{\BIBforeignlanguage}[2]{{%
\expandafter\ifx\csname l@#1\endcsname\relax
\typeout{** WARNING: IEEEtran.bst: No hyphenation pattern has been}%
\typeout{** loaded for the language `#1'. Using the pattern for}%
\typeout{** the default language instead.}%
\else
\language=\csname l@#1\endcsname
\fi
#2}}
\providecommand{\BIBdecl}{\relax}
\BIBdecl

\bibitem{recommendation2006vocabulary}
{ITUT Recommendation}, ``Vocabulary for performance and quality of service,''
  \emph{International Telecommunications Union—Radiocommunication (ITU-T),
  RITP: Geneva, Switzerland}, 2006.

\bibitem{pesq}
A.~W. Rix, J.~G. Beerends, M.~P. Hollier, and A.~P. Hekstra, ``Perceptual
  evaluation of speech quality {(PESQ)} - a new method for speech quality
  assessment of telephone networks and codecs,'' pp. 749--752, 2001.

\bibitem{moller2010comparison}
S.~M{\"o}ller, F.~Hinterleitner, T.~H. Falk, and T.~Polzehl, ``{Comparison of
  Approaches for Instrumentally Predicting the Quality of Text-to-speech
  Systems},'' in \emph{Proc. Interspeech}, 2010.

\bibitem{linear-reg-BC2012}
C.~R. Norrenbrock, F.~Hinterleitner, U.~Heute, and S.~M{\"o}ller, ``{Towards
  perceptual quality modeling of synthesized audiobooks-Blizzard Challenge
  2012},'' in \emph{Blizzard Challenge Workshop}, 2012.

\bibitem{norrenbrock2015quality}
------, ``{Quality Prediction of Synthesized Speech based on Perceptual Quality
  Dimensions},'' \emph{Speech Communication}, vol.~66, pp. 17--35, 2015.

\bibitem{Yoshimura2016}
T.~Yoshimura, G.~E. Henter, O.~Watts, M.~Wester, J.~Yamagishi, and K.~Tokuda,
  ``{A Hierarchical Predictor of Synthetic Speech Naturalness Using Neural
  Networks},'' in \emph{Interspeech}, 2016, pp. 342--346.

\bibitem{patton2016automos}
B.~Patton, Y.~Agiomyrgiannakis, M.~Terry, K.~Wilson, R.~A. Saurous, and
  D.~Sculley, ``{AutoMOS: Learning a Non-intrusive Assessor of
  Naturalness-of-speech},'' in \emph{Proc. NIPS End-to-end Learning for Speech
  and Audio Processing Workshop}, 2016.

\bibitem{mosnet}
C.-C. Lo, S.-W. Fu, W.-C. Huang, X.~Wang, J.~Yamagishi, Y.~Tsao, and H.-M.
  Wang, ``{MOSNet: Deep Learning-Based Objective Assessment for Voice
  Conversion},'' in \emph{Proc. Interspeech}, 2019, pp. 1541--1545.

\bibitem{siamese-mos}
G.~Mittag and S.~M{\"o}ller, ``{Deep Learning Based Assessment of Synthetic
  Speech Naturalness},'' in \emph{Proc. Interspeech}, 2020, pp. 1748--1752.

\bibitem{Choi2020}
Y.~Choi, Y.~Jung, and H.~Kim, ``{Deep MOS Predictor for Synthetic Speech using
  Cluster-Based Modeling},'' in \emph{Proc. Interspeech}, 2020, pp. 1743--1747.

\bibitem{williams2020comparison}
J.~Williams, J.~Rownicka, P.~Oplustil, and S.~King, ``Comparison of speech
  representations for automatic quality estimation in multi-speaker
  text-to-speech synthesis,'' \emph{Speaker Odyssey}, 2020.

\bibitem{mbnet}
Y.~Leng, X.~Tan, S.~Zhao, F.~Soong, X.-Y. Li, and T.~Qin, ``{MBNET: MOS
  Prediction for Synthesized Speech with Mean-Bias Network},'' in \emph{Proc.
  ICASSP}, 2021, pp. 391--395.

\bibitem{s3r-mos}
W.-C. Tseng, C.-Y. Huang, W.-T. Kao, Y.~Y. Lin, and H.-Y. Lee, ``{Utilizing
  Self-supervised Representations for MOS Prediction},'' in \emph{Proc.
  Interspeech}, 2021, pp. 2781--2785.

\bibitem{generalization-mos}
E.~Cooper, W.-C. Huang, T.~Toda, and J.~Yamagishi, ``{Generalization ability of
  MOS prediction networks},'' in \emph{Proc. ICASSP}, 2022.

\bibitem{BVCC}
E.~Cooper and J.~Yamagishi, ``How do voices from past speech synthesis
  challenges compare today?'' in \emph{Proc. SSW}, 2021, pp. 183--188.

\bibitem{blizzard2008}
V.~Karaiskos, S.~King, R.~A. Clark, and C.~Mayo, ``The {Blizzard Challenge}
  2008,'' in \emph{Proc. Blizzard Challenge Workshop}.\hskip 1em plus 0.5em
  minus 0.4em\relax Citeseer, 2008.

\bibitem{blizzard2009}
A.~W. Black, S.~King, and K.~Tokuda, ``The {Blizzard Challenge} 2009,'' in
  \emph{Proc. Blizzard Challenge}, 2009, pp. 1--24.

\bibitem{blizzard2010}
S.~King and V.~Karaiskos, ``The {Blizzard Challenge} 2010,'' 2010.

\bibitem{blizzard2011}
------, ``The {Blizzard Challenge} 2011,'' 2011.

\bibitem{blizzard2013}
------, ``The {Blizzard Challenge} 2013,'' 2013.

\bibitem{blizzard2016}
------, ``The {Blizzard Challenge} 2016,'' 2016.

\bibitem{vcc2016}
T.~Toda, L.-H. Chen, D.~Saito, F.~Villavicencio, M.~Wester, Z.~Wu, and
  J.~Yamagishi, ``{The Voice Conversion Challenge 2016},'' in \emph{Proc.
  Interspeech}, 2016, pp. 1632--1636.

\bibitem{vcc2018}
J.~Lorenzo-Trueba, J.~Yamagishi, T.~Toda, D.~Saito, F.~Villavicencio,
  T.~Kinnunen, and Z.~Ling, ``{The Voice Conversion Challenge 2018: Promoting
  Development of Parallel and Nonparallel Methods},'' in \emph{Proc. Odyssey},
  2018, pp. 195--202.

\bibitem{vcc2020}
Y.~Zhao, W.-C. Huang, X.~Tian, J.~Yamagishi, R.~K. Das, T.~Kinnunen, Z.~Ling,
  and T.~Toda, ``{Voice Conversion Challenge 2020 - Intra-lingual semi-parallel
  and cross-lingual voice conversion -},'' in \emph{Proc. Joint Workshop for
  the BC and VCC 2020}, 2020, pp. 80--98.

\bibitem{hayashi2019espnettts}
T.~Hayashi, R.~Yamamoto, K.~Inoue, T.~Yoshimura, S.~Watanabe, T.~Toda,
  K.~Takeda, Y.~Zhang, and X.~Tan, ``{{ESPnet-TTS}: Unified, Reproducible, and
  Integratable Open Source End-to-End Text-to-Speech Toolkit},'' 2020.

\bibitem{wu2019blizzard}
Z.~Wu, Z.~Xie, and S.~King, ``{The Blizzard Challenge 2019},'' in \emph{Proc.
  Blizzard Challenge Workshop}, vol. 2019, 2019.

\bibitem{mosanet}
R.~E. Zezario, S.-W. Fu, F.~Chen, C.-S. Fuh, H.-M. Wang, and Y.~Tsao, ``{Deep
  Learning-based Non-Intrusive Multi-Objective Speech Assessment Model with
  Cross-Domain Features},'' \emph{arXiv preprint arXiv:2111.02363}, 2021.

\bibitem{ldnet}
W.-C. Huang, E.~Cooper, J.~Yamagishi, and T.~Toda, ``{LDNet: Unified Listener
  Dependent Modeling in MOS Prediction for Synthetic Speech},'' in \emph{Proc.
  ICASSP}, 2022.

\bibitem{wav2vec2}
A.~Baevski, H.~Zhou, A.~Mohamed, and M.~Auli, ``wav2vec 2.0: A framework for
  self-supervised learning of speech representations,'' in \emph{NeurIPS 2020},
  2020.

\bibitem{LibriSpeech}
V.~Panayotov, G.~Chen, D.~Povey, and S.~Khudanpur, ``Librispeech: an asr corpus
  based on public domain audio books,'' in \emph{2015 IEEE international
  conference on acoustics, speech and signal processing (ICASSP)}.\hskip 1em
  plus 0.5em minus 0.4em\relax IEEE, 2015, pp. 5206--5210.

\bibitem{gururangan-etal-2020-dont}
S.~Gururangan, A.~Marasovi{\'c}, S.~Swayamdipta, K.~Lo, I.~Beltagy, D.~Downey,
  and N.~A. Smith, ``{Don{'}t Stop Pretraining: Adapt Language Models to
  Domains and Tasks},'' in \emph{Proc. ACL}, 2020, pp. 8342--8360.

\end{thebibliography}

\iftoggle{arxiv} {
    \appendix
    \onecolumn
    \section{Challenge Rules}
\label{appendix:rules}

We required all participants to make at least one submission to the CodaLab leaderboard during the training phase.  Although participants had access to the development set labels and could evaluate their models locally, we required submission to the leaderboard in order to ensure that participants were familiar with the submission format and procedures ahead of the more time-constrained evaluation phase.  Teams who did not make one submission to the leaderboard by the start of the evaluation phase were assumed to no longer be interested in participating and were removed from the challenge.

We also required participants to submit a system description for their team entry.  A template with example information filled in was provided. 

Participants were allowed to make use of external data for the challenge, and were required to specify information about any external data used in their system descriptions.  The exception is that participants were explicitly not allowed to gather and make use of data from Blizzard Challenges, Voice Conversion Challenges, or ESPnet-TTS published samples outside of what was provided for the challenge, because in doing so they may inadvertently use samples that appear in the test set for training.  Likewise, participants were not allowed to make use of pretrained models or other resources that used these datasets.

The development set was only allowed to be used for model selection and parameter tuning, and not as extra training data for the evaluation phase.

\section{List of participants}
\label{appendix:participants}
\begin{table*}[h]
\centering
\caption{List of participant affiliations in random order.}
\label{tab:teams}

\begin{tabular}{ l c c }
        \toprule
        Affiliation & Main track & OOD track \\
        \midrule
        Ajmide Media, China & Y & Y \\
        Budapest University of Technology and Economics, Hungary & Y & Y \\
        Bytedance AI-Lab, China & Y & Y \\
        Charles University, Prague, Czech Republic & Y & N \\
        Denso IT Laboratory, Japan & Y & Y \\
        Duke Kunshan University & Y & N \\
        Google; University College Dublin & Y & N \\
        Inner Mongolia University, China & Y & N \\
        Japan Advanced Institute of Science and Technology, Japan & Y & N \\
        National Taiwan University, Taiwan & Y & Y \\
        Netease, China & Y & Y \\
        NICT, Japan; Kyoto Univ., Japan; Kuaishou Inc., China & Y & Y \\
        Novosibirsk State University & N & Y \\
        Personal? & Y & Y \\
        Princeton University & Y & Y \\
        ReadSpeaker, The Netherlands & Y & N \\
        Sillwood Technologies, UK & Y & Y \\
        Technical University of Cluj-Napoca, Romania & Y & N \\
        The University of Tokyo, Japan & Y & Y \\
        Tsinghua University? & Y & Y \\
        University College Dublin, Ireland & Y & Y \\
        University of West Bohemia, Czech Republic & Y & Y \\
        \bottomrule

\end{tabular}
\end{table*}

\vfill\pagebreak

\section{Detailed results}
\label{appendix:detailed-results}

\subsection{Main track}
\begin{table*}[h!]
	\centering
	
	\caption{Main track evaluation results. For MSE, the smaller the better; for LCC, SRCC and KTAU, the larger the better. }

	\centering

	\begin{tabular}{ | c | c c c c | c c c c | }

        \hline
		\multirow{2}{*}[0pt]{ID} & \multicolumn{4}{c|}{Utterance-level} & \multicolumn{4}{c|}{System-level} \\
		\cline{2-9}

		& MSE & LCC & SRCC & KTAU & MSE & LCC & SRCC & KTAU \\

        \hline
        B01 & 0.277 & 0.869 & 0.869 & 0.690 & 0.148 & 0.924 & 0.921 & 0.769 \\ 
        B02 & 0.309 & 0.809 & 0.811 & 0.621 & 0.162 & 0.899 & 0.900 & 0.729 \\ 
        B03 & 0.334 & 0.787 & 0.785 & 0.599 & 0.178 & 0.873 & 0.873 & 0.691 \\
        \hline
        T01 & 0.208 & 0.875 & 0.878 & 0.704 & 0.122 & 0.916 & 0.915 & 0.767 \\ 
        T02 & 0.205 & 0.878 & 0.877 & 0.701 & 0.116 & 0.927 & 0.925 & 0.776 \\ 
        T04 & 0.351 & 0.805 & 0.806 & 0.619 & 0.194 & 0.890 & 0.890 & 0.719 \\ 
        T05 & 0.194 & 0.884 & 0.884 & 0.712 & 0.107 & 0.931 & 0.930 & 0.781 \\ 
        T06 & 0.238 & 0.861 & 0.860 & 0.686 & 0.143 & 0.923 & 0.922 & 0.773 \\ 
        T07 & 0.242 & 0.877 & 0.872 & 0.698 & 0.098 & 0.936 & 0.933 & 0.783 \\ 
        T08 & 0.263 & 0.838 & 0.838 & 0.652 & 0.142 & 0.908 & 0.911 & 0.747 \\ 
        T09 & 0.276 & 0.823 & 0.822 & 0.634 & 0.137 & 0.904 & 0.902 & 0.734 \\ 
        T10 & 0.197 & 0.884 & 0.884 & 0.712 & 0.106 & 0.933 & 0.931 & 0.785 \\ 
        T11 & 0.179 & 0.895 & 0.894 & 0.725 & 0.101 & 0.941 & 0.939 & 0.797 \\ 
        T12 & 0.171 & 0.895 & 0.893 & 0.725 & 0.095 & 0.936 & 0.933 & 0.782 \\ 
        T13 & 0.299 & 0.856 & 0.854 & 0.674 & 0.245 & 0.909 & 0.910 & 0.745 \\ 
        T14 & 0.414 & 0.748 & 0.747 & 0.562 & 0.249 & 0.837 & 0.834 & 0.649 \\ 
        T15 & 0.442 & 0.760 & 0.758 & 0.572 & 0.253 & 0.833 & 0.827 & 0.640 \\ 
        T16 & 0.313 & 0.835 & 0.829 & 0.647 & 0.142 & 0.911 & 0.906 & 0.745 \\ 
        T17 & 0.165 & 0.900 & 0.897 & 0.730 & 0.090 & 0.939 & 0.936 & 0.794 \\ 
        T18 & 0.213 & 0.882 & 0.884 & 0.712 & 0.130 & 0.925 & 0.929 & 0.783 \\ 
        T19 & 0.187 & 0.890 & 0.889 & 0.719 & 0.091 & 0.940 & 0.938 & 0.792 \\ 
        T20 & 0.197 & 0.889 & 0.890 & 0.722 & 0.108 & 0.936 & 0.935 & 0.790 \\ 
        T21 & 0.217 & 0.867 & 0.866 & 0.687 & 0.103 & 0.920 & 0.918 & 0.763 \\ 
        T22 & 0.220 & 0.867 & 0.864 & 0.687 & 0.134 & 0.912 & 0.911 & 0.749 \\

        \hline
	\end{tabular}
	\label{tab:raw-results-main}

\end{table*}
\begin{table*}[h!]
	\centering

	\caption{Main track rankings for each metric.}

	\centering

	\begin{tabular}{ | c | c c c c | c c c c | }

        \hline
		\multirow{2}{*}[0pt]{ID} & \multicolumn{4}{c|}{Utterance-level} & \multicolumn{4}{c|}{System-level} \\
		\cline{2-9}

		& MSE & LCC & SRCC & KTAU & MSE & LCC & SRCC & KTAU \\

        \hline
        B01 & 17 & 12 & 12 & 12 & 18 & 11 & 12 & 12 \\ 
        B02 & 19 & 20 & 20 & 20 & 19 & 20 & 20 & 20 \\ 
        B03 & 21 & 22 & 22 & 22 & 20 & 22 & 22 & 22 \\ 
        \hline
        T01 & 9 & 11 & 9 & 9 & 11 & 14 & 14 & 13 \\ 
        T02 & 8 & 9 & 10 & 10 & 10 & 9 & 10 & 10 \\ 
        T04 & 22 & 21 & 21 & 21 & 21 & 21 & 21 & 21 \\ 
        T05 & 5 & 6 & 7 & 7 & 8 & 8 & 8 & 9 \\ 
        T06 & 13 & 15 & 15 & 15 & 17 & 12 & 11 & 11 \\ 
        T07 & 14 & 10 & 11 & 11 & 4 & 4 & 5 & 6 \\ 
        T08 & 15 & 17 & 17 & 17 & 15 & 18 & 16 & 16 \\ 
        T09 & 16 & 19 & 19 & 19 & 14 & 19 & 19 & 19 \\ 
        T10 & 7 & 7 & 6 & 8 & 7 & 7 & 7 & 5 \\ 
        T11 & 3 & 2 & 2 & 3 & 5 & 1 & 1 & 1 \\ 
        T12 & 2 & 3 & 3 & 2 & 3 & 6 & 6 & 8 \\ 
        T13 & 18 & 16 & 16 & 16 & 22 & 17 & 17 & 18 \\ 
        T14 & 23 & 24 & 24 & 24 & 23 & 23 & 23 & 23 \\ 
        T15 & 24 & 23 & 23 & 23 & 24 & 24 & 24 & 24 \\ 
        T16 & 20 & 18 & 18 & 18 & 16 & 16 & 18 & 17 \\ 
        T17 & 1 & 1 & 1 & 1 & 1 & 3 & 3 & 2 \\ 
        T18 & 10 & 8 & 8 & 6 & 12 & 10 & 9 & 6 \\ 
        T19 & 4 & 4 & 5 & 5 & 2 & 2 & 2 & 3 \\ 
        T20 & 6 & 5 & 4 & 4 & 9 & 5 & 4 & 4 \\ 
        T21 & 11 & 14 & 13 & 14 & 6 & 13 & 13 & 14 \\ 
        T22 & 12 & 13 & 14 & 13 & 13 & 15 & 15 & 15 \\

        \hline
	\end{tabular}
	\label{tab:rankings-main}

\end{table*}

\vfill\pagebreak

\subsection{OOD track}
\begin{table*}[h!]
	\centering

	\caption{OOD track evaluation results. For MSE, the smaller the better; for LCC, SRCC and KTAU, the larger the better. }

	\centering

	\begin{tabular}{ | c | c c c c | c c c c | }

        \hline
		\multirow{2}{*}[0pt]{ID} & \multicolumn{4}{c|}{Utterance-level} & \multicolumn{4}{c|}{System-level} \\
		\cline{2-9}

		& MSE & LCC & SRCC & KTAU & MSE & LCC & SRCC & KTAU \\

        \hline
        B01 & 0.260 & 0.888 & 0.849 & 0.664 & 0.099 & 0.971 & 0.975 & 0.889 \\ 
        B02 & 0.284 & 0.853 & 0.806 & 0.616 & 0.071 & 0.967 & 0.954 & 0.846 \\ 
        B03 & 0.287 & 0.844 & 0.787 & 0.592 & 0.091 & 0.952 & 0.934 & 0.791 \\
        \hline
        T02 & 3.619 & 0.415 & 0.460 & 0.336 & 3.393 & 0.469 & 0.527 & 0.446 \\ 
        T03 & 0.216 & 0.896 & 0.870 & 0.683 & 0.066 & 0.976 & 0.929 & 0.803 \\ 
        T05 & 0.270 & 0.881 & 0.857 & 0.672 & 0.104 & 0.968 & 0.975 & 0.877 \\ 
        T06 & 0.719 & 0.874 & 0.848 & 0.662 & 0.538 & 0.957 & 0.970 & 0.871 \\ 
        T07 & 0.285 & 0.885 & 0.850 & 0.667 & 0.132 & 0.964 & 0.956 & 0.858 \\ 
        T10 & 0.220 & 0.884 & 0.852 & 0.664 & 0.085 & 0.955 & 0.930 & 0.778 \\ 
        T11 & 0.163 & 0.921 & 0.895 & 0.722 & 0.048 & 0.982 & 0.952 & 0.852 \\ 
        T12 & 0.254 & 0.894 & 0.863 & 0.674 & 0.107 & 0.968 & 0.936 & 0.809 \\ 
        T15 & 0.334 & 0.849 & 0.814 & 0.619 & 0.163 & 0.950 & 0.917 & 0.778 \\ 
        T17 & 0.162 & 0.917 & 0.893 & 0.716 & 0.030 & 0.988 & 0.979 & 0.908 \\ 
        T18 & 0.189 & 0.900 & 0.875 & 0.692 & 0.054 & 0.972 & 0.973 & 0.883 \\ 
        T19 & 0.210 & 0.888 & 0.859 & 0.676 & 0.070 & 0.960 & 0.969 & 0.871 \\ 
        T20 & 0.360 & 0.858 & 0.837 & 0.649 & 0.141 & 0.964 & 0.956 & 0.846 \\ 
        T21 & 2.516 & 0.440 & 0.429 & 0.305 & 2.261 & 0.491 & 0.516 & 0.385 \\ 
        T22 & 0.261 & 0.885 & 0.834 & 0.646 & 0.107 & 0.957 & 0.904 & 0.785 \\

        \hline
	\end{tabular}
	\label{tab:raw-results-ood}

\end{table*}
\begin{table*}[h!]
	\centering

	\caption{OOD track rankings for each metric.}

	\centering

	\begin{tabular}{ | c | c c c c | c c c c | }

        \hline
		\multirow{2}{*}[0pt]{ID} & \multicolumn{4}{c|}{Utterance-level} & \multicolumn{4}{c|}{System-level} \\
		\cline{2-9}

		& MSE & LCC & SRCC & KTAU & MSE & LCC & SRCC & KTAU \\

        \hline
        B01 & 8 & 6 & 10 & 10 & 9 & 5 & 2 & 2 \\ 
        B02 & 11 & 14 & 15 & 15 & 6 & 8 & 9 & 9 \\ 
        B03 & 13 & 16 & 16 & 16 & 8 & 15 & 12 & 13 \\ 
        \hline
        T02 & 18 & 18 & 17 & 17 & 18 & 18 & 17 & 17 \\ 
        T03 & 5 & 4 & 4 & 4 & 4 & 3 & 14 & 12 \\ 
        T05 & 10 & 11 & 7 & 7 & 10 & 6 & 2 & 4 \\ 
        T06 & 16 & 12 & 11 & 11 & 16 & 12 & 5 & 5 \\ 
        T07 & 12 & 9 & 9 & 8 & 13 & 9 & 7 & 7 \\ 
        T10 & 6 & 10 & 8 & 9 & 7 & 14 & 13 & 15 \\ 
        T11 & 2 & 1 & 1 & 1 & 2 & 2 & 10 & 8 \\ 
        T12 & 7 & 5 & 5 & 6 & 12 & 7 & 11 & 11 \\ 
        T15 & 14 & 15 & 14 & 14 & 15 & 16 & 15 & 15 \\ 
        T17 & 1 & 2 & 2 & 2 & 1 & 1 & 1 & 1 \\ 
        T18 & 3 & 3 & 3 & 3 & 3 & 4 & 4 & 3 \\ 
        T19 & 4 & 7 & 6 & 5 & 5 & 11 & 6 & 5 \\ 
        T20 & 15 & 13 & 12 & 12 & 14 & 10 & 7 & 9 \\ 
        T21 & 17 & 17 & 18 & 18 & 17 & 17 & 18 & 18 \\ 
        T22 & 9 & 8 & 13 & 13 & 11 & 13 & 16 & 14 \\

        \hline
	\end{tabular}
	\label{tab:rankings-ood}

\end{table*}

\vfill\pagebreak

\section{Correlation analysis between metrics}
\label{appendix:metrics-correlations-all}

\begin{table}[h]
\centering
\caption{Linear correlation coefficients between different metrics, using the main track results.}
\label{tab:metrics-correlations-all}

\begin{tabular}{ | c c | c c c c | c c c c | }
    \hline
    & & \multicolumn{4}{c|}{Utterance-level} & \multicolumn{4}{c|}{System-level} \\
    & & MSE & LCC & SRCC & KTAU & MSE & LCC & SRCC & KTAU \\
    \hline \hline
    \multirow{4}{*}[0pt]{\makecell{Utterance-\\level}} & MSE & 1.00 & -.955 & -.958 & -.955 & .906 & -.931 & -.932 & -.942 \\
    & LCC & - & 1.00 & .999 & .997 & -.838 & .964 & .959 & .974 \\
    & SRCC & - & - & 1.00 & .997 & -.835 & .962 & .959 & .974 \\
    & KTAU & - & - & - & 1.00 & -.829 & .949 & .944 & .965\\
    \hline
    \multirow{4}{*}[0pt]{\makecell{System-\\level}} & MSE & - & - & - & - & 1.00 & -.875 & -.862 & -.870 \\
    & LCC & - & - & - & - & - & 1.00 & .997 & .994\\
    & SRCC & - & - & - & - & - & - & 1.00 & .994\\
    & KTAU & - & - & - & - & - & - & - & 1.00\\

\hline
\end{tabular}
\end{table}

\section{Factors investigated to explain systems that were difficult to predict}

\begin{figure}[h!] 
\centering
\includegraphics[width=1.\linewidth]{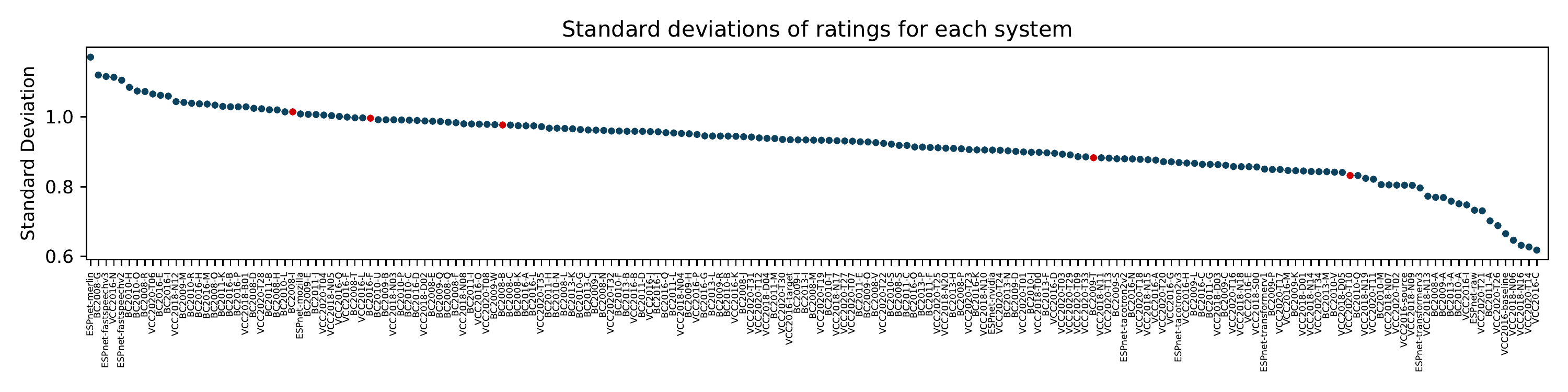}
\vspace{-6mm}
\caption{Systems and their standard deviations.  The five most difficult systems to predict (red) span the range of standard deviations, indicating that a high standard deviation in a system's scores does not explain its difficulty for MOS predictors.}
\label{fig:emd}

\end{figure}

\begin{figure}[h!] 
\centering
\includegraphics[width=1.\linewidth]{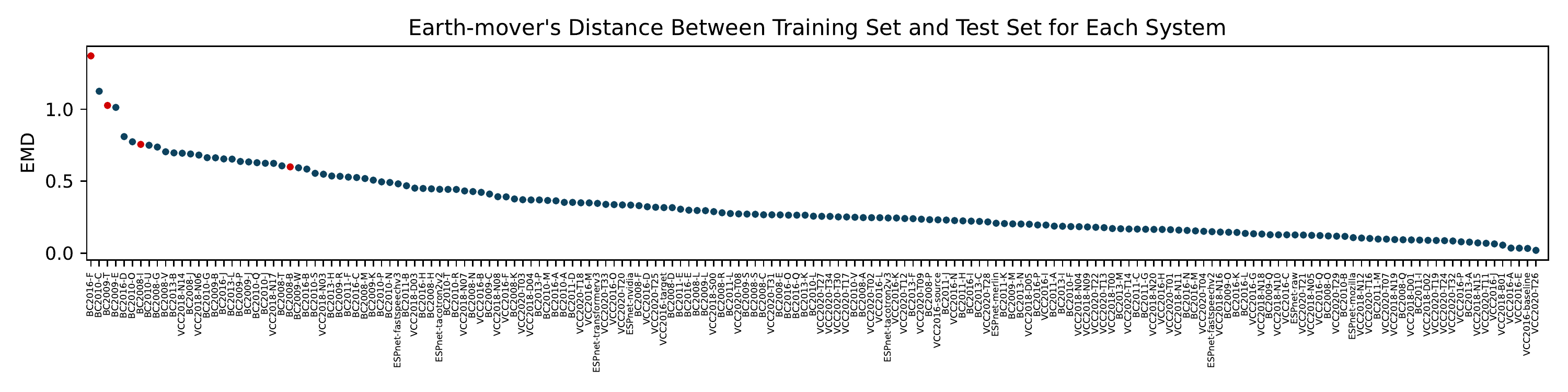}
\vspace{-6mm}
\caption{Difference in distributions between training and test data.  The four most difficult systems to predict (red) are in the top EMD range of this figure (left side), indicating that large differences in the distributions of the training and test data contribute to prediction difficulty.}
\label{fig:emd}

\end{figure}

\begin{figure}[h!] 
\centering
\includegraphics[width=1.\linewidth]{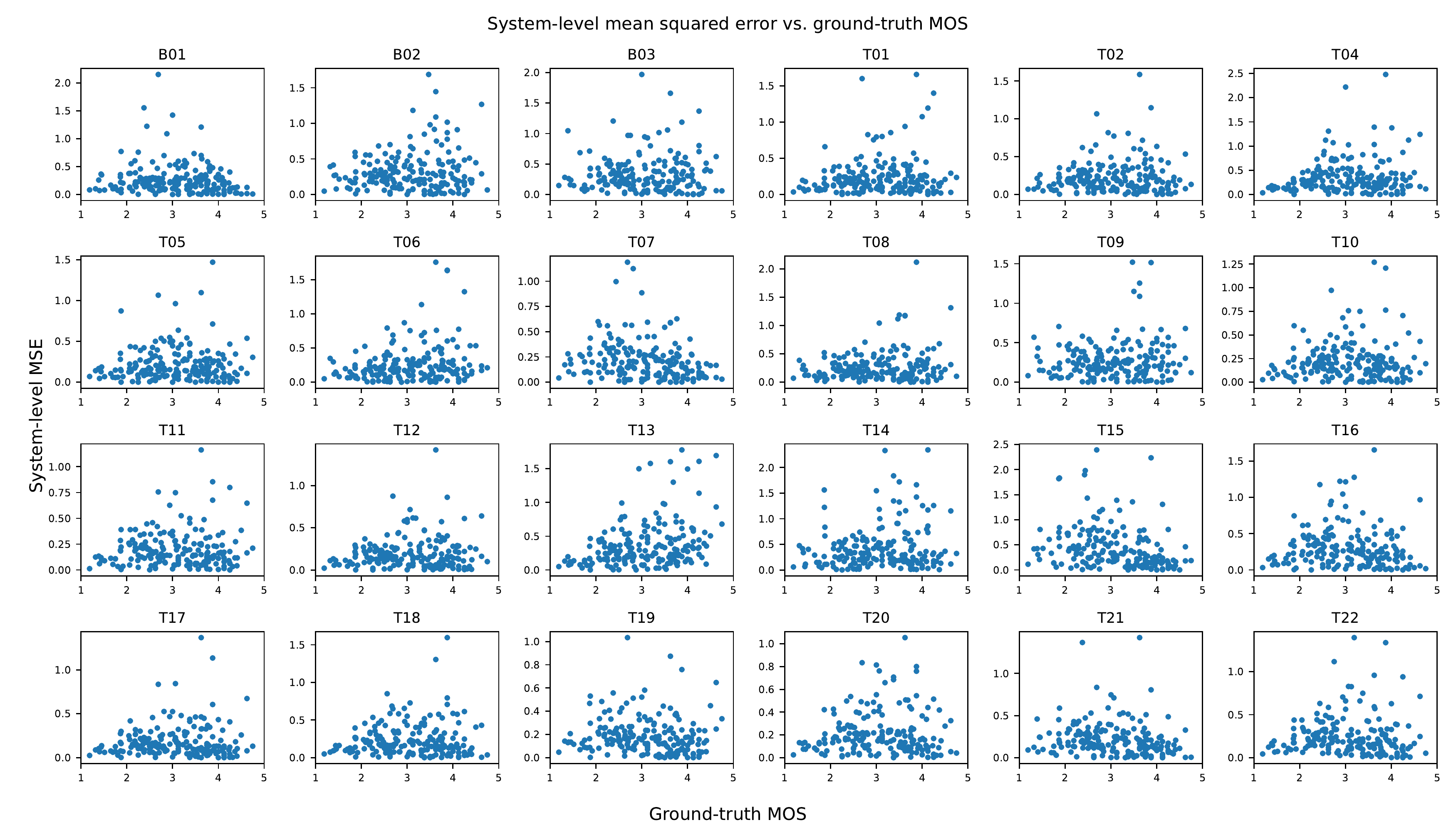}
\vspace{-6mm}
\caption{System-level mean squared error vs. ground-truth system-level MOS.  All teams had low errors for low-scoring systems.  Higher errors tend to appear for middle- and high-scoring systems.}
\label{fig:emd}

\end{figure}
}

\end{document}